# Measurement of Absolute Single and Double Electron Capture Cross Sections for $O^{6+}$ Ions Collision with $CO_2$, $CH_4$, $H_2$ and $N_2$

J. Han[1], L. Wei[1], B. Wang[1], B. Ren[1], W. Yu[1], Y. Zhang[2], Y. Zou[1], L. Chen[1], J. Xiao[1], and B. Wei[1]

[1] Institute of Modern Physics, Key Laboratory of Nuclear Physics and Ion-Beam Application (MOE), Fudan University, Shanghai 200433, China; xiao_jun@fudan.edu.cn, brwei@fudan.edu.cn

[2] School of Mathematics, Physics and Information Engineering, Jiaxing University, Jiaxing 314001, China

**Abstract**

The absolute electron capture cross sections for single and double charge exchanges between the highly charged ions $O^{6+}$ and $CO_2$, $CH_4$, $H_2$, $N_2$, the dominant collision processes in the solar wind, have been measured in the energy from 7 keV•q (2.63 keV/u) to 52 keV•q (19.5 keV/u). These measurements were carried out in the new experimental instrument set up at Fudan University, and the error of cross sections for single and double charge exchanges at the 1σ confidence level are about 11% and 16%, respectively. Limited agreement is achieved with single electron capture results calculated by the classical over-barrier model. These cross sections data are useful for simulation ion-neutral processes in astrophysical environments and to improve the present theoretical model of fundamental atomic processes.

## 1. Introduction

After the ROSAT satellite (Lisse et al. 1996) observed X-ray emissions from comet Hyakutake that far exceeded expectations, it is believed that the phenomenon can be explained by charge exchange (CE) theory between solar wind (SW) heavy ions and comet material (Cravens 1997). Since then, observation and model analysis on the X-ray spectrum of comets (Wegmann et al. 1998; Bodewits et al. 2007; Snios, et al. 2016), Mars (Dennerl et al. 2006), Jupiter (Branduardi-Raymont et al. 2007) and Jovian planets (Hui et al. 2009), and outer heliosphere, such as in supernova remnants (Katsuda et al. 2011), starburst galaxies (Liu, Mao & Wang 2011) were successively carried out, further confirming that the CE process has an important influence on the cosmic X-ray. In modeling CE emission, the availability of atomic data for CE is often insufficient, thus hindering the completeness and validity of present model. Cumbee et al. (2016) and Mullen et al. (2016) respectively used theoretical methods such as classical trajectory Monte Carlo (CTMC) and Landau-Zener (LZ) methods to calculate the cross sections of relative CE processes as the physical data input of the analysis model. Greenwood et al. (2000), Čadež et al. (2003) and Moradmand et al. (2018) measured the CE cross sections for highly charged ions (HCIs) in SW colliding with gas targets such as $H_2$, $H_2O$, $CO_2$, etc.

As the important heavy-ion constituents in the SW, oxygen in the form of $O^{6+}$ ions, retaining only two orbital electrons is the most abundant in the SW (Neugebauer et al. 2000; Cravens 2002). The result of $^{16}O/^{18}O$ density ratio from the observations of isotopes in the SW (Collier et al. 1998; Wimmer-Schweingruber et al. 1999) indicate that $^{16}O$ accounts for the majority compared to $^{18}O$. Mawhorter et al. (2007) and Machacek et al. (2014; 2015) measured the cross sections of single and double CEs for $O^{6+}$ colliding with $H_2O$, $CO_2$ and other gas targets at the energy of 3.5 keV•q and 7 keV•q for discussing the astrophysical environments and processes. In this article, considering the energy range of SW and the field-aligned potential acceleration of heavy ions in Jovian magnetosphere, single and double electron capture (EC) cross sections from 7 keV•q (2.63 keV/u) to 52 keV•q (19.5 keV/u) for $^{16}O^{6+}$ colliding with abundant cometary and planetary species $CO_2$, $CH_4$, $H_2$, $N_2$ are accurately measured by a new experimental instrument set up at Fudan University, which fills the blank in the data of EC cross sections for HCIs and provides experimental basis for the verification of theoretical calculations (Gao et al. 2017; Zhang et al. 2020) in the energy range of 10 -100keV/u. To provide theoretical explanation, calculations of the single EC cross sections are estimated by classical over-barrier model (OBM).

## 2. Experiment method

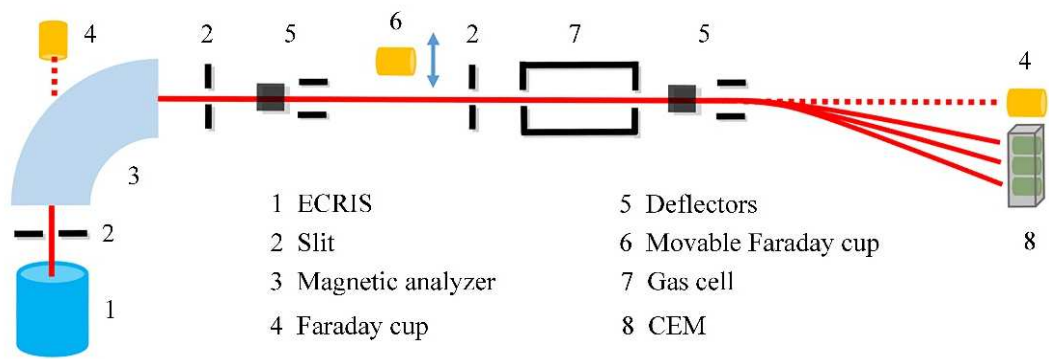

**Figure 1.** Schematic drawing of the experimental apparatus

The experiments were performed on a HCIs platform based on electron cyclotron resonance (ECR) ion source constructed at Fudan University (Zhang et al. 2018). An all permanent magnet 14.5 GHz ECR ion source, delivered by Pantechik, is installed on a high voltage platform (150 kV) to provide large currents of multiply charged ion beams. Figure 1 shows the main features of the experimental arrangement in a schematic way. Ion beams of corresponding elements with different charge states are generated by injecting gases into ECR ion source. After the acceleration, the charge state of highly charged ions is mass-to-charge selected by a 90° magnet analyzer. Then the ions beam is guided into the gas cell which is full of target gas, and the background vacuum is better than $10^{-8}$ torr without sample gas. The pressure in the gas cell is measured with a MKS Baratron absolute capacitance diaphragm gauge in real time which was kept below 20 mTorr. By passing through the voltage deflection plate, scattering ions in different charge states due to the charge transfer process are separated and introduced into the suitable channel electron multiplier (CEM) for detection.

The adjustment and monitoring of ion beams are mainly through slits, deflection plates, and Faraday cups. The shape and intensity of the ion beam is optimized by adjusting multiple different slits. Correcting the direction of the beam is achieved by applying a deflection voltage. Three Faraday cups at different positions including a movable one, monitor beam current several times to ensure the position and intensity of the projectile beam. The important criterion for identifying whether the quality of beam current is suitable for the experiment is that $N^0_{q-1}$: $N^0_{total}$ ($N^0_{total}$ is the total number of beam particles, $N_{q-1}$ is the number of particles recorded as having charge $q$ -1) of the ion beam should be less than 1% (Knudsen et al. 1981) without target gas in the gas cell.

The experimental measurement method of the cross section is realized by the mathematical form of the growth rate method (Tawara & Russek 1973). Due to there are various charge transfer processes between HCIs and gas targets and multiple charge transfer processes between secondary particles, the cross sections of different charge transfer processes have intricate mathematical relationships, thus it is difficult to obtain value of the CE cross sections simply and directly. For this reason, by maintaining the gas pressure at an extremely low state, not only the influence of the process of converting other possible charge states to the target charge state, but also the mathematical difficulties caused by multiple collisions can be neglected. In this single collision condition, the measurement and calculation of the cross section of the charge transfer process is simplified to the study on the dependence of the initial growth rate of different charged ion beams on the gas target pressure (Tawara & Russek 1973). Accordingly, the formula for calculating the cross section can be transformed into the following mathematical form:

$$\sigma_{q,q-1} = \left(\frac{dF_{q-1}}{d\pi}\right)_{\pi=0}, \tag{1}$$

$$F_{q-1}(\pi) = \sigma_{q,q-1}\pi. \tag{2}$$

$F_{q-1}$ is the ratio of the ions with charge of $q$-1 to the total, and the target thickness $\pi$ is the number of target gas molecules in the volume whose area is 1 cm², given by

$$\pi = nl = \frac{Pl}{kT}, \tag{3}$$

where $n$ is the gas molecule number density, $l$ is the effective length of gas cell, $k$ is Boltzmann constant, and $T$ is the temperature of gas cell in Kelvin. The growth rate curve of $O^{6+}$ colliding with gas He is shown in Figure 2. When $\pi$ reaches $4\times10^{14}$ cm⁻² (P = 0.27mTorr), the linear dependence of $F_{q-1}$ as a function of $\pi$ verifies the approximation of single collisions, showing no effect of multiple collisions on the cross sections at such a low pressure.

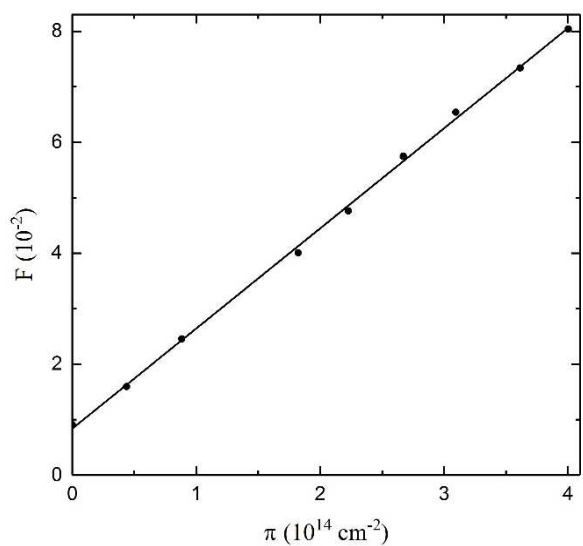

**Figure 2.** Example of a growth rate curve 2.63 keV/u $O^{6+}$ in collision with He that shows the fraction of $O^{5+}$ as a function of the target thickness. The single EC cross section was derived from the slope of the linear fit.

*2.1. Measurement error*

The uncertainties in the measured cross sections result primarily from the following factors: (1) the measurements of temperature and pressure, (2) the effective path length of the gas cell, (3) the detection efficiency, (4) incident ion current stability and (5) data statistics.

The ion beam enters the gas cell through the 1 mm beam inlet at the front end, passes through the target gas and undergoes the charge transfer process, finally exits from the 3 mm outlet at the rear end, the length of the gas cell is 60.8 mm. After gas injection, the gas molecules flow will reach dynamic equilibrium quickly, so that the value of the pressure gauge is approximately the same as the real pressure in the center of the gas cell. However, due to the influence of the gas flow at the inlet and outlet, the uncertainty of effective path length is estimated to be 6.6% (Toburen et al. 1968). The temperature of the gas cell (23.6℃), which is far different with the working temperature of pressure gauge (49℃) cause 1% error, and the fluctuation of temperature cause 0.1%.

Ions from the gas cell obtain a certain vertical displacement and velocity through the deflection field of the applied appropriate voltage, enter the field-free drift zone, and are finally received by the CEM detector. In particular, vertical displacement $y_1$ is

$$y_1 = \frac{l_1^2 q_2 V_2}{4 d q_1 V_1}, \tag{4}$$

and vertical velocity $v_y$ is

$$v_y = \frac{l_1 q_2 V_2}{d\sqrt{2 m q_1 V_1}}. \tag{5}$$

The vertical offset on final arrival $y_2$ is

$$y_2 = \frac{l_1 l_2 q_2 V_2}{2 d q_1 V_1}, \tag{6}$$

and relative displacement of different charge states $\Delta y$ is derived by the following formula

$$\Delta y = \frac{(2 l_2 + l_1) l_1 \Delta q_2 V_2}{4 d q_1 V_1} = F(\Delta q_2). \tag{7}$$

where $l_1$, $l_2$ are the length of deflection zone and drift zone, $q_1$, $q_2$ are the charge state before and after EC process, $V_1$ is ionic acceleration voltage, $V_2$ is deflection voltage, $m$ is ionic mass and $d$ is the distance between deflection plates. The calculation result shows that the relative position on the detector of the adjacent charge states $q$ and $q$-1 does not vary with the charge $q$. Therefore, the position of the detectors corresponding to different charges is determined by calculation, which shows the full collection of ions through experimental test. In addition, there is two orders of magnitude difference between the pressure of the scatter pipe (deflection zone and drift zone) and the gas cell after gas injection, which causes 2% error.

The detection system is mainly composed of five parts: lens, converter plate, electrode, CEM and shielding shell. Secondary electrons generated from metal plate are extracted through the lens and electrodes, and enter the electron multiplier tube to be detected. Rinn et al. (1982) systematically discussed the similar CEM detection system and found that the efficiency of detection is mainly affected by the counting rate. During the measurement experiment, the general counting rate ($s^{-1}$) of $q$, $q$-1, $q$-2 is controlled at about $10^4$, $10^3$, $10^2$, so that the single and double EC cross sections are 5.1% and 5.7% deviation. For the detector itself, the uncertainty of the detection efficiency is estimated to be 6% between CEMs for measuring different charged ions. In addition, the efficiency of CEM is constant within a mean deviation of 0.015 (Rinn et al. 1982) within an area of a 10-mm diameter, which ensures the full collection of beam whose spot diameter is about 3 mm.

For statistical error, it mainly comes from the counting statistics of the CEM. Due to double EC cross section of which the value is generally small, less count will bring relatively large errors. Thus, the summary of the sources and magnitudes of the experimental errors is given in Table 1. The total error is calculated by the error transfer formula.

**Table 1**
Individual and total experimental errors at the 1σ confidence level

| Source of error | Error at the 1σ confidence level (%) |
|---|---|
| Error in Temperature | |
|     Temperature difference between chamber and gauge | 1 |
|     Temperature fluctuation | 0.1 |
| Error in Pressure | |
|     Accuracy of pressure gauge | 0.25 |
|     Pressure fluctuation | 3 |
| Error in gas cell length | 6.6 |
| Error in count of CEM | |
|     Uncertainty of detection efficiency | 8 |
|     Effect of background vacuum | 2 |
| Error in incident ion current stability | |
|     $\sigma_{q,q-1}$ | 2 |
|     $\sigma_{q,q-2}$ | 10 |
| Data statistics | |
|     $\sigma_{q,q-1}$ | 2 |
|     $\sigma_{q,q-2}$ | 6 |
| Total error | |
|     $\sigma_{q,q-1}$ | 11 |
|     $\sigma_{q,q-2}$ | 16 |

## 3. Results and discussion

Table 2 and Figure 3 show the measurement results, the total cross sections of single and double EC in $^{16}O^{6+}$ colliding with $CO_2$, $CH_4$, $H_2$ and $N_2$ from 2.63 keV/u to 19.5 keV/u. Since autoionization and photon emission from ions in excited states following the charge exchange process whose time scale is larger than the time interval between the CE process and the scattered ions detected by the detector which is up to sub-microsecond, the experimental data mentioned in this paper are the absolute total cross sections of single and double EC, which means a single transfer and an autoionizing double transfer could contribute to the total single-capture cross section $\sigma_{q,q-1}$ and double transfer, single-autoionizing triple capture, and double-autoionizing quadruple capture contribute to the double-capture cross section $\sigma_{q,q-2}$ (Machacek et al. 2014). It is shown that the single and double cross sections of each target remain consistent within the energy range. The cross sections of collision with $CO_2$ at ion energy 2.63 keV/u are basically consistent with the data measured by Mawhorter et al. (2007), which are 4.83±0.37, 1.36±0.12, respectively. For hydrogen gas target, the measured results of single EC from Dijkkamp et al. (1985) ($^{16}O^{6+}$) and Machacek et al. (2014) ($^{18}O^{6+}$) are all presented in Figure 4 with our data which extend the collision energy region.

**Table 2**

Absolute Measurements of Single and Double EC Cross Sections for $^{16}O^{6+}$ Colliding with $CO_2$, $CH_4$, $H_2$ and $N_2$

| keV/u | process | Target | | | |
|---|---|---|---|---|---|
| | | $CO_2$ | $CH_4$ | $H_2$ | $N_2$ |
| 2.63 | $\sigma_{q,q-1}$ | 5.23 ± 0.58 | 5.03 ± 0.55 | 4.47 ± 0.49 | 5.18 ± 0.57 |
| | $\sigma_{q,q-2}$ | 1.48 ± 0.24 | 1.96 ± 0.31 | 0.31 ± 0.05 | 0.95 ± 0.15 |
| 4.50 | $\sigma_{q,q-1}$ | 4.90 ± 0.54 | 4.71 ± 0.52 | 4.47 ± 0.49 | 4.62 ± 0.51 |
| | $\sigma_{q,q-2}$ | 1.53 ± 0.24 | 1.87 ± 0.30 | 0.38 ± 0.06 | 1.01 ± 0.16 |
| 6.00 | $\sigma_{q,q-1}$ | 5.01 ± 0.55 | 5.44 ± 0.60 | 4.64 ± 0.51 | 4.74 ± 0.52 |
| | $\sigma_{q,q-2}$ | 1.60 ± 0.26 | 2.07 ± 0.33 | 0.36 ± 0.06 | 1.23 ± 0.20 |
| 9.00 | $\sigma_{q,q-1}$ | 4.51 ± 0.50 | 5.32 ± 0.59 | 4.61 ± 0.51 | 4.21 ± 0.46 |
| | $\sigma_{q,q-2}$ | 1.39 ± 0.22 | 2.09 ± 0.33 | 0.37 ± 0.06 | 1.05 ± 0.17 |
| 12.0 | $\sigma_{q,q-1}$ | 4.65 ± 0.51 | 5.11 ± 0.56 | 4.43 ± 0.49 | 4.33 ± 0.48 |
| | $\sigma_{q,q-2}$ | 1.43 ± 0.23 | 1.94 ± 0.31 | 0.41 ± 0.07 | 1.09 ± 0.17 |
| 15.0 | $\sigma_{q,q-1}$ | 4.18 ± 0.46 | 5.26 ± 0.58 | 4.26 ± 0.47 | 3.96 ± 0.44 |
| | $\sigma_{q,q-2}$ | 1.39 ± 0.22 | 2.03 ± 0.33 | 0.45 ± 0.07 | 1.09 ± 0.17 |
| 19.5 | $\sigma_{q,q-1}$ | 4.32 ± 0.48 | 4.84 ± 0.53 | 4.32 ± 0.48 | 4.26 ± 0.47 |
| | $\sigma_{q,q-2}$ | 1.42 ± 0.23 | 1.93 ± 0.31 | 0.42 ± 0.07 | 1.17 ± 0.19 |

**Note.** All cross sections are in units of $10^{-15}$ cm$^2$ and errors are given at 1σ level.

**Table 3**

Comparison of Measured and Calculated Single EC Cross Sections for $CO_2$, $CH_4$, $H_2$ and $N_2$

| Target Gas | $O^{6+}$ | | |
|---|---|---|---|
| | n | $\sigma_{q,q-1}$ Measured | $\sigma_{q,q-1}$ Calculated |
| $CO_2$ | 4 | 4.69 | 5.75 |
| $CH_4$ | 4 | 5.10 | 5.03 |
| $H_2$ | 4 | 4.46 | 7.07 |
| $N_2$ | 4 | 4.47 | 7.21 |

**Note.** Projectile energies and measured cross sections are the averages of the values in Table 2. All cross sections are in units of $10^{-15}$ cm$^2$. The value n is the $O^{6+}$ quantum number into which the electron is captured.

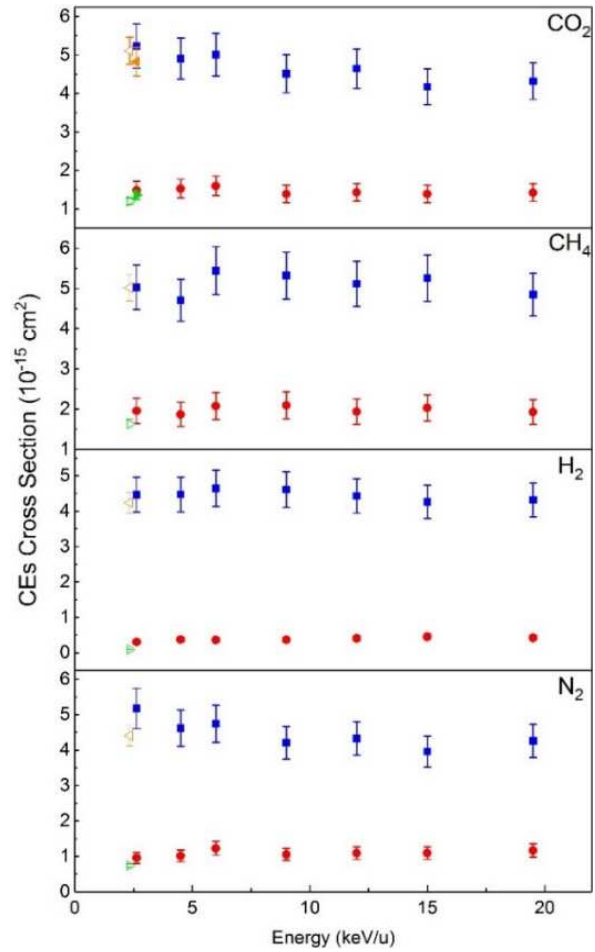

**Figure 3.** Measured absolute single (filled squares) and double (filled circles) total EC cross sections for $^{16}O^{6+}$ colliding with $CO_2$, $CH_4$, $H_2$ and $N_2$. Errors are given at 1σ level. For comparison, absolute single (◀ and ◁), double (▶ and ▷) total EC cross sections for $^{16}O^{6+}$ and $^{18}O^{6+}$ measured by Mawhorter et al. (2007) and Machacek et al. (2014; 2015) are also shown in the picture.

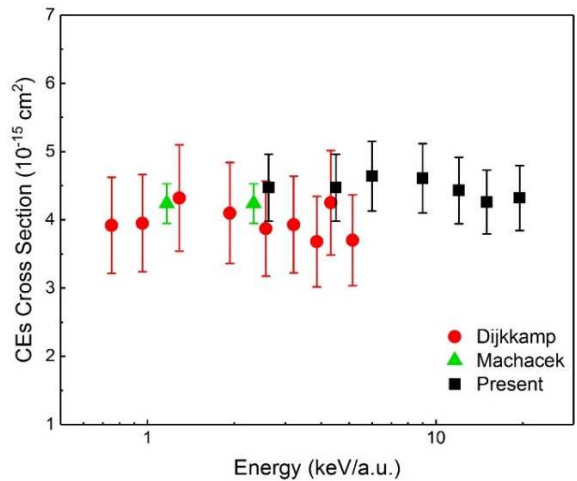

**Figure 4.** Single EC cross sections as a function of impact energy for $O^{6+}$ colliding with $H_2$. The experimental results are from Dijkkamp et al. (1985) (●), Machacek et al. (2014) (▲), and the present measurement (■).

As the molecular target changes, the single EC cross sections only vary slightly. Specifically, measured single EC cross sections vary by only 4% on the average ($4.87\pm0.22\times10^{-15}$ cm$^2$) for the molecular targets. The physical reason that the single EC cross section does not change greatly with the different types of gas molecular targets is that the EC process is mainly affected by the first ionization energy. For example, the first ionization energies of $CO_2$, $CH_4$, $H_2$ and $N_2$ are 13.78 eV, 12.61 eV, 15.43 eV and 15.58 eV, respectively, showing 9% deviation from the average value (14.35 eV). Therefore, since the little variation in energies by which the first electron to be removed, the population—both in relative number and dominantly populated $n\ell$ levels—is similar for single EC of $O^{6+}$ colliding with any of these molecules (Machacek et al. 2015).

To provide some theoretical explanation, calculations of the single EC cross sections were estimated by OBM (Mann et al. 1981). The principal quantum number $n$ is predicted to be the largest integer satisfying the inequality

$$n \leq q[2|I_p|(1+\frac{q-1}{2\sqrt{q+1}})]^{-\frac{1}{2}}. \qquad (8)$$

Here, $I_p$ is the ionization potential (in atomic units) of the target molecule, and $q=6$ is the $O^{q+}$ charge state. The crossing distance $Rx$ for collision systems is

$$R_x = \frac{q-1}{\frac{q^2}{2n^2}-|I_p|}. \qquad (9)$$

The single EC cross section is given by $\pi R_x^2$, whose results are displayed in Table 3. For comparison, the average of the measured single EC cross sections at the average energy value of 9.80 keV/u is also shown in Table 3. For the same incident ion charge state and close ionization potential, capture state $n$ of different targets are all equal to 4, so that the different calculation results of single EC cross sections are only resulted from the difference in ionization potential. Nevertheless, as a relatively simple theoretical model, OBM lacks consideration of the influence of many other physical factors such as ion incident energy, which makes the deviation of the mean theoretical calculation from the average of measurements up to 39%. It indicates that OBM cannot be widely applied to accurate theoretical calculation of single EC cross sections.

## 4. Conclusions

An experimental device for measuring the absolute cross section of EC has been introduced in detail. The total error corresponding to the single and double EC at the 1σ confidence level are given at 11% and 16%, respectively. Total absolute cross sections of single and double EC in the energy range 2.63-19.5 keV/u $^{16}O^{6+}$ colliding with $CO_2$, $CH_4$, $H_2$ and $N_2$ have been measured, which provides an experimental basis for judging the correctness of the theoretical calculations and fills in the blank of the EC cross sections data in the corresponding energy region. The agreement is found with earlier measurements of Mawhorter et al. (2007) with data for $CO_2$. In this energy range, the single capture cross sections remain basically the same and vary slightly with the target gas for the similar ionization potential. Calculations of the single EC cross sections have also been estimated by OBM. Part of the theoretical calculations is in good agreement with the experimental data. These results are useful for understanding the effects of the solar wind and basic atomic processes.

J. Han and L. Wei contributed equally to this work. This work was supported by the National Natural Science Foundation of China under Grant No. U1832201 and 11674067, the National Key Research and Development Program of China under Grant No. 2017YFA0402300.